\documentclass[10pt]{article}
\usepackage{euscript,amssymb,amsmath}
\usepackage[dvips]{graphicx}
\catcode`\@=11

\newcommand{\be}[1]{\begin{equation}\label{#1}}
\newcommand{\ee}{\end{equation}}
\newcommand{\ba}[1]{\begin{eqnarray}\label{#1}}
\newcommand{\ea}{\end{eqnarray}}
\newcommand{\rf}[1]{(\ref{#1})}
\newcommand{\nn}{\nonumber}

\newcommand{\diag}{\mbox{\rm diag}\,}

\begin{document}

\title{Brouwer's problem on a heavy particle in a rotating vessel: \\ wave propagation, ion traps, and rotor dynamics }
\maketitle
\begin{center}
{\large Oleg N. Kirillov } \\ Magneto-Hydrodynamics Division (FWSH), Helmholtz-Zentrum Dresden-Rossendorf\\
P.O. Box 510119, D-01314 Dresden, Germany\\
(e-mail:~~o.kirillov@fzd.de)~~%olkirillov@mail.ru%) %Tel: +49 6151 16 6828, Fax: +49 6151 16 4125)
\end{center}
\date{}

\begin{abstract}
In 1918 Brouwer considered stability of a heavy particle in a rotating vessel. This was the first demonstration of a rotating saddle trap which is a mechanical analogue for quadrupole particle traps of Penning and Paul. We revisit this pioneering work in order to uncover its intriguing connections with classical rotor dynamics and fluid dynamics, stability theory of Hamiltonian and non-conservative systems as well as with the modern works on crystal optics and atomic physics. In particular, we find that the boundary of the stability domain of the undamped Brouwer's problem possesses the Swallowtail singularity corresponding to the quadruple zero eigenvalue. In the presence of dissipative and non-conservative positional forces
there is a couple of Whitney umbrellas on the boundary of the asymptotic stability domain. The handles of the umbrellas form a set where all eigenvalues of the system are pure imaginary despite the presence of dissipative and non-conservative positional forces.
\end{abstract}

\begin{flushleft}
{%MSC: 34D20; 34D10; 34C23\\
}
\end{flushleft}

\section{Introduction}

%Diederik Korteweg his teacher.%

In 1918 Luitzen Egbertus (Bertus) Jan Brouwer (1881--1966) --- a founder of modern topology who established, for example, the topological invariance of dimension and the fixpoint theorem --- considered a problem on stability of a heavy particle in a rotating vessel \cite{Br18}. The work written in Dutch was almost forgotten until in 1976 Bottema drew attention to it \cite{B76}. However, even those later publications did not manage to popularize substantially the Brouwer's work, which is remarkable in many respects.

\begin{figure}
    \begin{center}
    \includegraphics[angle=0, width=0.75\textwidth]{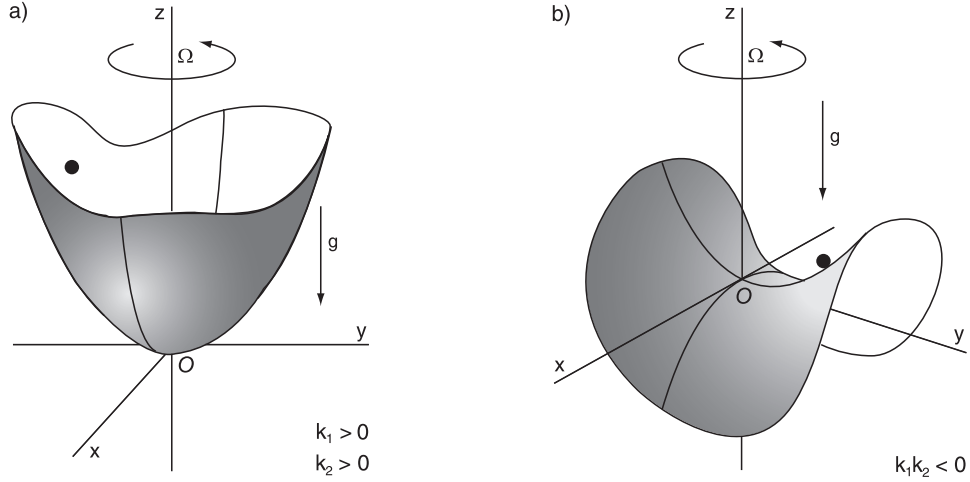}
    \end{center}
    \caption{A heavy particle in a rotating vessel \cite{Br18}. }
    \label{fig0}
    \end{figure}

In Brouwer's setting, a heavy particle of unit mass is moving on a surface that has a horizontal tangential plane at some point $O$ and rotates with a constant angular velocity $\Omega$ around a vertical axis through $O$, Fig.~\ref{fig0}. A rectangular coordinate system $(x, y, z)$ with the origin at $O$ and the $(x,z)$- and $(y,z)$-plane of which coincide with the principal normal sections of the surface at $O$, rotates with the surface's angular velocity around the $z$-axis  in a counter-clockwise direction, Fig.~\ref{fig0}.

Taking into account that a potential function of the particle is $V=gz$ and a kinetic energy is $T=\frac{1}{2}[(\dot x - y \Omega)^2+(\dot y + x \Omega)^2 +\dot z^2]$, Brouwer derives the non-linear equations of motion both in the frictionless case and in the case of Coulomb friction between the particle and the surface. The frictionless equations linearized around the equilibrium position $O$, have the form \cite{Br18}
\ba{i1}
\ddot x - 2\Omega \dot y+\left(k_1-\Omega^2 \right)x&=&0,\nn\\
\ddot y + 2\Omega \dot x+\left(k_2-\Omega^2 \right)y&=&0,
\ea
where dot indicates time differentiation, $k_{1,2}=g/r_{1,2}$, and $r_{1,2}$ are the radii of curvature of the intersection of the surface with the $(x, y)$- and $(x, z)$-plane, respectively, $r_1\ne r_2$. A merit of Brouwer is that he not only allowed the radii of curvature to be non-equal, which was already the state-of-the-art of his time, but he also did not impose restrictions on the sign of $k_1$ and $k_2$ and considered both concave (Fig.~\ref{fig0}(a)) and saddle-like (Fig.~\ref{fig0}(b)) surfaces, ahead of his contemporaries.

Analyzing the characteristic equation of the system \rf{i1}
\be{i2}
\lambda^4+(k_1+k_2+2\Omega^2)\lambda^2+(k_1-\Omega^2)(k_2-\Omega^2)=0,
\ee
Brouwer writes down the conditions for marginal stability of the particle
\ba{i3}
(k_1-\Omega^2)(k_2-\Omega^2)&>&0,\nn\\
k_1+k_2+2\Omega^2&>&0,\nn\\
(k_1-k_2)^2+8\Omega^2(k_1+k_2)&>&0.
\ea

It is instructive to plot the stability domain \rf{i3} in the $(k_1,k_2,\Omega)$-space. We restrict our study by the non-negative values of $\Omega$ because the stability domain is symmetric with respect to the plane $\Omega=0$.

\begin{figure}
    \begin{center}
    \includegraphics[angle=0, width=0.7\textwidth]{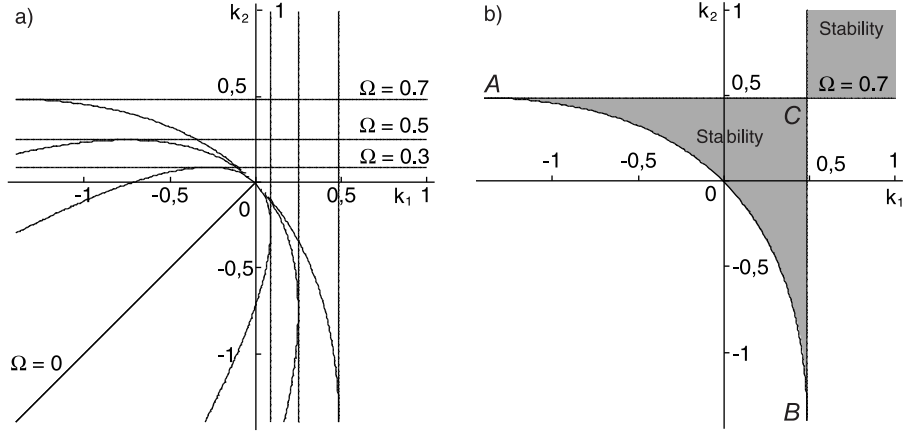}
    \end{center}
    \caption{(a) Contours of the Whitney umbrella and two parabolic cylinders \rf{i3} in the $(k_1, k_2)$-plane for $\Omega=0, 0.3, 0.5, 0.7$ and (b) the domain of the marginal stability in the same plane for $\Omega=0.7$. }
    \label{fig1}
    \end{figure}

First we note that the parabolic cylinders $k_1=\Omega^2$ and $k_2=\Omega^2$ intersect along the parabola that lies in the plane $k_1=k_2$. This parabola smoothly touches the parabolic cylinder $k_1+k_2+2\Omega^2=0$ as well as the Whitney umbrella $k_1+k_2=-(k_1-k_2)^2/(8\Omega^2)$ at the origin. The Whitney umbrella smoothly touches both of the parabolic cylinders $k_1=\Omega^2$ and $k_2=\Omega^2$ along spacial curves that project into the lines $k_2=-3k_1$ and $k_1=-3k_2$ in the $(k_1, k_2)$-plane, see Fig.~\ref{fig1}(a).

In Fig.~\ref{fig1}(a), the contours of the Whitney umbrella and of the two parabolic cylinders are shown for various values of $\Omega$. One sees that the contours of the Whitney umbrella are parabolas that shrink into a fragment of a single line $k_1=k_2$ that lies in the third quadrant of the $(k_1,k_2)$-plane. The contours of the parabolic cylinders are straight lines parallel to the coordinate axes. The parabolas touch the straight lines, the latter intersect each other along the line $k_1=k_2$. Therefore, the stability domains have the form of a curvilinear triangle $ABC$ with cuspidal point singularities at the two corners $A$ and $B$ and with the transversal intersection singularity at the third one marked as the point $C$ in Fig.~\ref{fig1}(b).  At the singular point $C$ the triangular stability domain ({\em stability for large angular velocities} according to Brouwer) is joint with the infinite stability domain given by the inequalities $k_1>\Omega^2$, $k_2>\Omega^2$ ({\em stability for small angular velocities} according to Brouwer). The latter part of the stability diagram was found already in 1869 by Rankine \cite{Ran} whereas the triangle $ABC$ was reconstructed by the joint efforts of F\"oppl (1895) \cite{Foe95}, von K\'arm\'an (1910) \cite{K10}, Prandtl (1918) \cite{Pr18}, Brouwer (1918) \cite{ Br18}, and Jeffcott (1919) \cite{J19}. Note that similar planar stability domains appear in a recent study of levitation of a rotating body carrying a point electrical charge in the field of a fixed point charge of the same sign \cite{DNF10}, see also \cite{Br10}.

A remarkable result of Brouwer is that for $k_1k_2<0$, i.e. when the surface is a saddle, the particle can be stabilized by rotation of the surface. Brouwer concludes that `as long as in $O$ the principal
curvature, concave in an upward direction, is less than three times the one concave
in a downward direction, there are rotation velocities for which the motion of the
particle on the rotating saddle yields formal stability' \cite{Br18}.
Cuspidal points $A$ and $B$ on the stability boundary clearly seen in Fig.~\ref{fig1}(b) evidence that such a stabilization occurs for a rather non-trivial choice of the main curvatures of the rotating saddle.

    \begin{figure}
    \begin{center}
    \includegraphics[angle=0, width=0.4\textwidth]{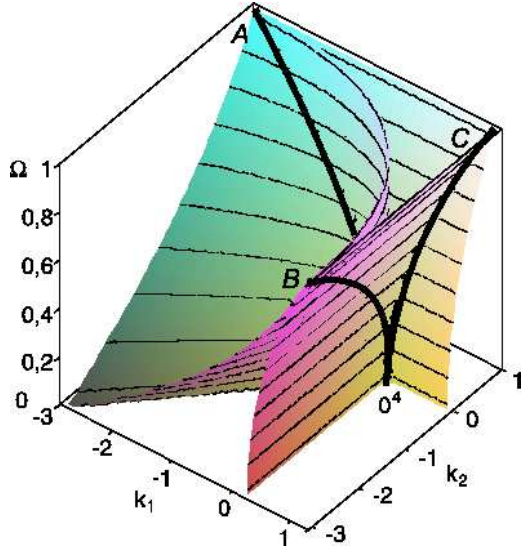}
    \end{center}
    \caption{The Swallowtail singularity $(0^4, \cite{Ga82})$ at the origin on the stability boundary of the Brouwer's problem on a heavy particle in a rotating vessel illustrating peculiarities of gyroscopic stabilization. }
    \label{fig2}
    \end{figure}

We notice however, that while Brouwer \cite{Br18} neither plotted the stability domain nor discussed its geometry,
Bottema \cite{B76} improved Brouwer's result and even produced a planar stability diagram which however does not detect the singularities because of the unsatisfactory choice of parameters. Therefore, we substantially refine the analysis performed by Brouwer and Bottema by finding the singularities of the planar stability boundary for the system \rf{i1}.
Moreover, we discover that in the $(k_1,k_2,\Omega)$-space the full stability domain given by inequalities \rf{i3} is bounded by a surface that has the {\em Swallowtail} singularity at the origin, see Fig.~\ref{fig2}. Bold lines in Fig.~\ref{fig2} indicate two cuspidal edges $(A,B)$ and one regular intersection $(C)$ on the stability boundary. Thus, the part of the stability domain that corresponds to stability for large angular velocities is inside the {\em spike} of the singular surface whereas the part of the stability domain that corresponds to stability for small angular velocities occupies the remaining bigger portion of it that has only a regular edge singularity. The tangent cone to the stability domain at the tip of the spike consists of the semi-axis $\Omega \ge 0$, Fig.~\ref{fig2}. Similar structure of the stability domain was found in the model of a two-link rotating shaft with four degrees of freedom \cite{Br10,SM03}.

At the origin, the conditions $k_1=k_2=0$ and $\Omega=0$ yield quadruple zero eigenvalue $\lambda=0$ with the Jordan block.
Galin \cite{Ga82} had proven that this spectral degeneracy $(0^4)$ corresponds to the Swallowtail singularity of the stability boundary of a Hamiltonian system and has codimension 3.  The Swallowtail singularity generically appears in the Brouwer's problem that is described by the three parameter Hamiltonian system \rf{i1}.
We note that in the works \cite{Br10,SM03} the Swallowtail singularity had been found on the stability boundary of a two-link rotating shaft that has twice as many degrees of freedom. The Brouwer's problem on a heavy particle in a rotating vessel appears to be the simplest physically meaningful system that possesses this singularity and as we will show further in the text, this fundamental paradigmatic model has deep connections to many important effects of classical and modern physics.

\section{Manifestations in physics}

The singular surface bounding the stability domain of the Brouwer's heavy particle in a rotating vessel perfectly
symbolizes the variety of applications to which it is connected. Almost literally at every corner of this surface lives
a physical phenomenon.

\subsection{Helical quadrupole magnetic focussing systems}

Already in 1914, in his correspondence to Georg Hamel, Brouwer discussed the possibility of experimental verification of stabilization of a heavy ball on a rotating saddle \cite{Br18}. Nowadays such mechanical demonstrations are available as teaching laboratory experiments \cite{THB2002}. However, the prospects for the first non-trivial physical application of this effect had arisen as early as 1936 when Penning proposed using the quadrupole electric field to confine the charged particles \cite{P36, BG86}.

In the RF-electric-quadrupole Paul trap invented in 1953 \cite{Pa53, P1990}, a saddle-shaped field
created to trap a charged ion is not rotating about the ion in the center. The Paul-trap potential can only `flap' the field up and down, which yields two decoupled Mathieu equations describing the motions of a single ion in the trap. Nevertheless, comparing the rotating saddle trap and the Paul trap, Shapiro \cite{S96,Sh01} and Thompson et al.~\cite{THB2002} demonstrated that the former mimics most of the characteristics of the Paul trap such as stability and instability regions, micromotion and secular oscillation frequency \cite{THB2002}.
Despite the Brouwer's problem is not only a mechanical analogy to the Paul trap, the equations of the two models are substantially different. Brouwer's system is autonomous, which greatly simplifies its solution. An example of a natural appearance of the
Brouwer's equations comes from accelerator physics, where they originate in a theory of focussing by a helical magnetic quadrupole lens \cite{Pe70}.

Indeed, according to Pearce \cite{Pe70} the linearized equations of motion in the laboratory $(x, y, z)$-frame of a particle of momentum $p$ and charge $e$ in the helical quadrupole magnetic field that rotates with distance along the $z$-axis, completing one rotation in an axial distance $\lambda$, in the assumption that $x, y \ll \lambda$, are
\ba{hq1}
\ddot x&=&a(x\cos z + y \sin z),\nn \\
\ddot y&=&a(x\sin z - y \cos z),
\ea
where dot stands for the derivative $d/dz$ and
\be{hq2}
a=- \frac{\lambda^2 k^2}{16 \pi^2},\quad k^2=\frac{eG}{pc},
\ee
and $c$ is the speed of light. The downstream distance $z$ is measured in units of $\lambda/(4 \pi)$.
At $z=0$, the components of the magnetic field in the laboratory frame are $B_x=Gy$, $By=Gx$, $B_z=0$ as for the conventional quadrupole.

By coordinate transformation to a frame $(X, Y, z)$, which rotates with the mechanical
rotation of the helix:
\ba{hq3}
X&=&~~x\cos\frac{z}{2}+y\sin\frac{z}{2},\nn \\
Y&=&-x\sin\frac{z}{2}+y\cos\frac{z}{2},
\ea
we arrive at the autonomous system of equations that is equivalent to \rf{hq1}
\be{hq4}
\ddot {\bf v}+{\bf J}\dot {\bf v}+({\bf K}+({\bf J}/2)^2){\bf v}=0,
\ee
where
\be{hq5}
{\bf v}=\left(
          \begin{array}{c}
            X \\
            Y \\
          \end{array}
        \right),\quad
{\bf J}=\left(
          \begin{array}{cc}
            0 & -1 \\
            1 & 0 \\
          \end{array}
        \right),\quad
{\bf K}=\left(
          \begin{array}{cc}
            -a & 0 \\
           0 & a \\
          \end{array}
        \right).
\ee
We note that \rf{hq4} is a particular case of the Brouwer's rotating vessel equations \rf{i1} with $\Omega=1/2$ and $k_2=-k_1=a$ that corresponds to a rotating saddle.

\subsection{Propagation of polarized light within the cholesteric liquid crystal}
Following Marathay \cite{Ma71} we consider the state of polarization of a plane wave as it propagates along the helical axis within the nonmagnetic cholesteric liquid crystal. Assume that $z$-axis of a right-handed laboratory frame is parallel to the helical axis. The molecular planes of the structure are parallel to the $(x,y)$-plane of the laboratory frame. The principal axes of each molecular plane are gradually rotated as one proceeds along $z$-direction. Define the pitch $p$ as the minimum distance between the two planes whose principal axes are parallel and introduce the parameter $\alpha=2\pi/p$. Restricting ourselves by the plane waves that propagate normally to the crystal planes, i.e. along the helical axis, we find from the Maxwell's equations that
\be{cr1}
\partial^2_z {\bf E} = \frac{\varepsilon(z)}{c^2}  \partial^2_t {\bf E},\quad {\bf E}=\left(
                                                                                       \begin{array}{c}
                                                                                         E_x \\
                                                                                         E_y \\
                                                                                       \end{array}
                                                                                     \right),
\ee
where ${\bf E}={\bf E}(z,t)$ is the Jones vector \cite{Ma71}. The  $2\times2$ matrix $\varepsilon(z)$
describes the dielectric properties of the layer at $z$ in the laboratory frame and for a right-handed structure in terms
of the principal dielectric constants $\varepsilon_{11}^0$ and $\varepsilon_{22}^0$ it can be expressed as
\be{cr2}
\varepsilon(z) = {\bf R}(\alpha z) {\bf K} {\bf R}^{-1}(\alpha z),
\ee
where ${\bf K}=\diag(\varepsilon_{11}^0, \varepsilon_{22}^0 )$ and {\bf R} is the rotation matrix
\be{cr3}
{\bf R}=\left(
  \begin{array}{rr}
   \cos(\alpha z) & -\sin(\alpha z) \\
    \sin(\alpha z) & ~\cos(\alpha z)\\
  \end{array}
\right).
\ee

Looking for a time-harmonic solution of Eq.~\rf{cr1} of the form
${\bf E}(z,t) ={\bf F}(z) exp(-i\omega t)$ and transforming to a space-rotating coordinate system by
\be{cr4}
{\bf f}(z) = {\bf R}(-\alpha z){\bf F}(z),
\ee
we find that in the space rotating coordinate system the dynamics
of the state of polarization ${\bf f}$ is governed by the autonomous equation
\be{cr5}
\ddot{\bf f}+2\alpha {\bf J} \dot {\bf f}+(k^2 {\bf K}+(\alpha {\bf J})^2){\bf f}=0,
\ee
where $k^2=\omega^2/c^2$ and the dot stands for the derivative $d/dz$. We again obtained the Brouwer's equations \rf{i1}, with $\Omega=\alpha$, $k_1=k^2\varepsilon_{11}^0$ and $k_2=k^2\varepsilon_{22}^0$.

\subsection{Benjamin-Feir instability of traveling waves}
The modulations of Stokes waves in deep water are described by the Non-Linear Schr\"odinger equation (NLS)
\be{bf00}
iA_t+\alpha A_{xx}+\gamma |A|^2A=0,
\ee
where $i=\sqrt{-1}$, $A$ is the envelope of the wave carrier, $\alpha$ and $\gamma$ are
positive real numbers and the modulations
are restricted to one space dimension $x$, for details see recent work by Bridges and Dias \cite{BD07}.

Assuming $A=u_1+iu_2$, linearizing the non-linear partial differential equation \rf{bf00} about the basic traveling wave solution with the amplitude ${\bf u}_0=(u_1^0,u_2^0)$, spacial wavenumber $k$ and the frequency $\omega$ and then expanding the periodic in $x$ solution with the period $\sigma $ of the linearized problem into Fourier series, we find that the $\sigma$-dependent modes decouple into four-dimensional
subspaces for each harmonic with the number $n$, so that for $n=1$ we get
\ba{bf0}
{\bf J}\dot {\bf v}+2\alpha k \sigma {\bf J}{\bf w}-\alpha \sigma^2 {\bf v}+2\gamma {\bf u}_0{\bf u}_0^T{\bf v}&=&0,\nn\\
{\bf J}\dot {\bf w}-2\alpha k \sigma {\bf J}{\bf v}-\alpha \sigma^2 {\bf w}+2\gamma {\bf u}_0{\bf u}_0^T{\bf w}&=&0,
\ea
where dot indicates time differentiation, the matrix $\bf J$ is defined in \rf{hq5}, the dyad ${\bf u}_0{\bf u}_0^T$ is a $2 \times 2$ symmetric matrix and $\omega^2=\alpha k^2-\|{\bf u}_0\|^2$.
The NLS model \rf{bf00} has a Benjamin-Feir instability: considering all other parameters fixed, with the increase of the wave amplitude ${\bf u}_0$ at some threshold
value two eigenvalues of the linear stability problem \rf{bf0} experience the {\em Krein collision} \cite{M98}, and these two modes have negative and positive energy, respectively \cite{BD07}.

Differentiating the first of the equations \rf{bf0} once, substituting into the result the expression for ${\bf J}\dot{\bf w}$ that follows from the second of the equations \rf{bf0}, then expressing ${\bf w}$ from the first of the equations \rf{bf0} and finally multiplying the result by $\bf J$ from the left we obtain the equation that describes the time evolution of  $\bf v$
\be{bf1}
\ddot {\bf v}-2{\bf J}{\bf B}\dot{\bf v}+(4\alpha^2k^2\sigma^2+({\bf JB})^2){\bf v}=0,
\ee
where ${\bf B}=2\gamma{\bf u}_0{\bf u}_0^T-\alpha\sigma^2{\bf I}$, and $\bf I$ is the unit matrix. The vector $\bf w$ satisfies the same equation. The matrix $\bf JB$ is non-symmetric and thus can be decomposed into a symmetric and skew-symmetric parts
\be{bf2}
{\bf J}{\bf B}=q{\bf J}+\gamma{\bf D},
\ee
where $q=\sigma^2\alpha -\gamma \|{\bf u}_0 \|^2$ and ${\bf D}={\bf u}_0{\bf u}_0^T{\bf J}-{\bf J}{\bf u}_0{\bf u}_0^T$.

With this notation we reduce the equation \rf{bf1} to the standard for the gyroscopic systems form
\be{bf3}
\ddot {\bf v}+2q{\bf J}\dot {\bf v}+2\gamma{\bf D}\dot{\bf v} +({\bf P}+(q{\bf J})^2){\bf v}=0,
\ee
where ${\bf P}=(4\alpha^2k^2\sigma^2+\gamma^2\|{\bf u}_0 \|^4){\bf I}$.

Equation \rf{bf3} is the damped Brouwer's equation \rf{i1} with $\Omega=q$ and $k_1=k_2=4\alpha^2k^2\sigma^2+\gamma^2\|{\bf u}_0 \|^4$. The `damping' matrix $\bf D$ is symmetric and traceless, i.e. it is indefinite with the eigenvalues  $\mu_{1,2}=\pm \| {\bf u}_0\|^2$.
Therefore we can give a new interpretation to the Benjamin-Feir instability and treat it as the \emph{destabilization of a gyroscopic system by the indefinite damping} \cite{KP09}. Remarkably, this destabilization mechanism is known in rotor dynamics, where the indefinite damping matrix results from the falling friction characteristic that typically occurs in the problems of acoustics of friction such as the squealing brake or the singing wine glass \cite{KP09,Ki08,Ki09b}. In \cite{Ki09b} destabilization of the Brouwer-type equations by the indefinite damping was considered in detail.

\subsection{Stability of deformable rotors}

With the equal positive coefficients $k_1=k_2>0$, the Brouwer's equations \rf{i1} coincide with that of the idealized model of the classical rotor dynamics problem of
shaft whirl by F\"oppl (1895) \cite{Foe95}, von K\'arm\'an (1910) \cite{K10}, and Jeffcott (1919) \cite{J19}, written in the rotating $(x,y)$-frame.

\begin{figure}
    \begin{center}
    \includegraphics[angle=0, width=\textwidth]{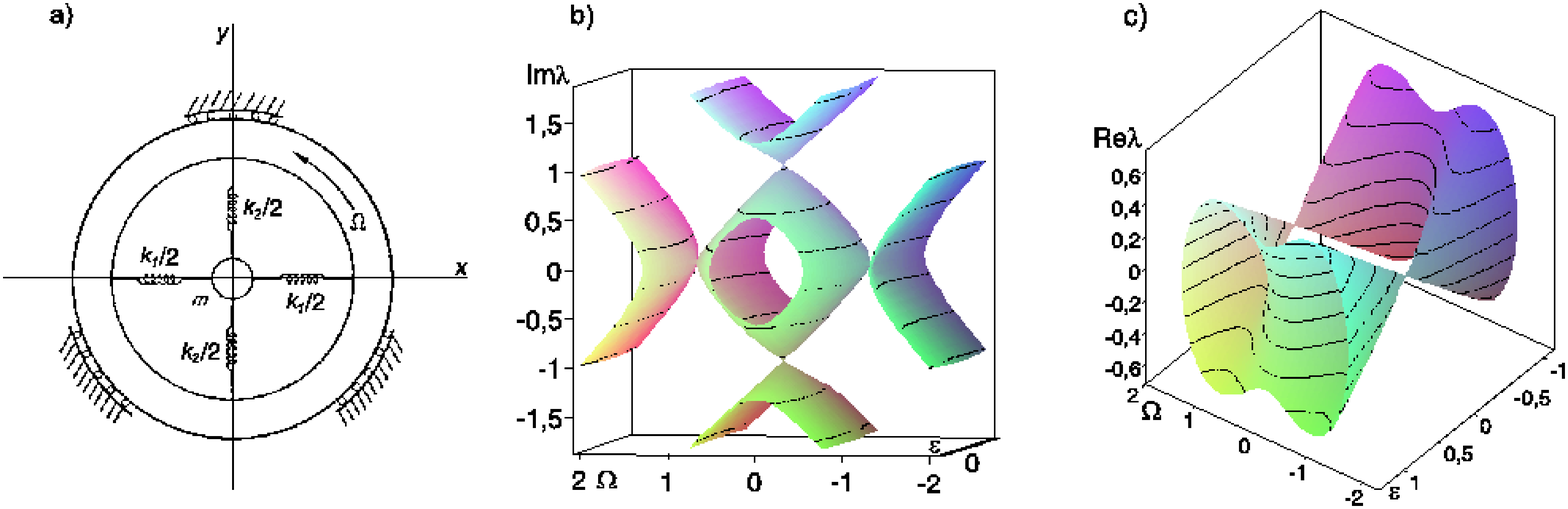}
    \end{center}
    \caption{(a) A Prandtl-Smith model of whirling shaft \cite{Cr70, Cr95}: A point mass $m=1$ suspended by the springs from the frame that rotates with the angular velocity $\Omega$. (b) Whirling frequencies of the Prandtl-Smith rotor with $k_1=1$ as a function of $\Omega$ and $\varepsilon$ form a singular surface with four conical points in the plane $\varepsilon=0$. (c) Growth rates of the Prandtl-Smith rotor showing instability due to stiffness detuning near the critical speeds $\Omega=\pm 1$.}
    \label{fig3}
    \end{figure}

Indeed, a deformable shaft carrying, e.g., a turbine wheel, and rotating with the angular speed $\Omega$ about its axis of symmetry can be modeled as a planar oscillator on a rotating plate \cite{HBW}, i.e. as a unit mass point that is suspended symmetrically by massless springs with the effective stiffness coefficients $k_1=k_2=k$ from the frame that rotates with the angular velocity $\Omega$ \cite{Cr70, Cr95}, see Fig.~\ref{fig3}(a).

For the symmetrical F\"oppl-von K\'arm\'an-Jeffcott rotor the transverse bending modes occur in pairs with equal
natural frequencies $\omega_n=\sqrt{k}$ for vibrations in orthogonal diametral planes. When the
vibrations of such a pair are combined with equal amplitudes and a quarter
period phase difference, the result is a circularly polarized vibration --- a clockwise- or a counter-clockwise circular whirling motion with the whirl rate $\omega_n$ \cite{Cr70, Cr95}. This phenomenon is related to the Doppler splitting of the doublet modes into the forward and backward traveling waves, which propagate along the circumferential direction of a rotating elastic solid of revolution discovered by Bryan in 1890 \cite{B1890}.

According to the Brouwer's stability conditions \rf{i3} the symmetrical F\"oppl-von K\'arm\'an-Jeffcott rotor
is marginally stable at any speed $\Omega$. Indeed, the part of the plane $k_1=k_2$ corresponding to $k_{1,2}>0$ belongs to the domain of marginal stability, as is seen in the stability diagrams of Fig.~\ref{fig1}(b) and Fig.~\ref{fig2}.

The symmetry of the F\"oppl-von K\'arm\'an-Jeffcott rotor is, however, a latent source of its instability.
Destabilization can be caused already by constraining the mass point to vibrate along a rotating
diameter. For example, if a guide rail is installed along the $x$-axis so that $y$ is constrained to
vanish identically, equation \rf{i1} reduces to the 1869 model by Rankine \cite{Ran}
\be{ro1}
\ddot x + (k_1 - \Omega^2)x = 0,
\ee
which predicts unbounded growth for $x$ as soon as $\Omega^2>k_1$, i.e. as soon as the centrifugal
field overpowers the elastic field \cite{Cr70, Cr95}. In the diagram of Fig.~\ref{fig1}(b) Rankine's instability threshold (critical rotating speed) bounds the infinite stability domain with the corner at the point $C$ showing that the constraint could also be achieved by tending one of the stiffness coefficients to infinity. Note that recently Baillieul and Levi motivated by stabilization of satellites with flexible parts, studied the dynamical effects of imposing constraints on the relative motions of component parts in a rotating mechanical system or structure and in particular in the Brouwer's equations \cite{BL91}.

Although after successful demonstration by De Laval of a well-balanced gas turbine running stably at supercritical, i.e. unstable by Rankine, speeds in 1889, the model of Rankine \cite{Ran} has been recognized as inadequate \cite{Cr70, Cr95}, we see that its instability threshold is just a part of the singular stability boundary of the Brouwer's equations shown in Fig.~\ref{fig2}.

Soon after the success of the symmetric F\"oppl-von K\'arm\'an-Jeffcott model of De Laval's rotor, Prandtl (1918) \cite{Pr18}
noted that if the elastic or inertia properties of a rotor are not symmetric about
the axis of rotation  there may be speed ranges where the rotor whirl is unstable.
He pointed out the analogy between
a rotor whose shaft had unequal principal stiffness coefficients $(k_1\ne k_2)$ and a pendulum mounted asymmetrically on a turntable.
The pendulum was unstable when the turntable rotation
rate $\Omega$ was between the pendulum natural frequencies in the soft and stiff directions. Similar instability due to inertia
asymmetry was discussed by Smith in 1933 \cite{S33}. For positive coefficients $k_1\ne k_2$, the Brouwer's equations \rf{i1} is exactly the undamped model by Prandtl and Smith.

Let us assume in the equation \rf{i1} that  $k_1=1+\varepsilon$ and $k_2=1-\varepsilon$  \cite{Cr70, Cr95}. Calculating the imaginary
parts of the roots of the characteristic equation \rf{i2} we find the implicit equation for the whirling frequencies in the
Prandtl-Smith model
\be{ro2}
(({\rm Im}\lambda)^2-1-\Omega^2)^2-4\Omega^2=\varepsilon^2.
\ee
In the $(\Omega,\varepsilon,{\rm Im}\lambda)$-space the whirling frequencies lie on a singular surface that has four conical points, Fig.~\ref{fig3}(b). The slice of this surface by the plane $\varepsilon=0$ shows four straight lines that intersect each other at the conical points at $\Omega=0$ and at the critical speeds $\Omega=\pm1$.
These eigencurves constitute the Campbell diagram \cite{C1924} of the ideal F\"oppl-von K\'arm\'an-Jeffcott rotor. Stiffness modification with the variation of $\varepsilon$ corresponds to the eigencurves
in a slice of the eigenvalue surface by the plane that departs of the conical singularities by a distance $\varepsilon\ne 0$.
In the vicinity of $\Omega=0$ the eigencurves demonstrate avoided crossings while near the critical speeds $\Omega=\pm1$ the eigenvalues are real with zero frequency. As is seen in Fig.~\ref{fig3}(c), the real eigenvalues (growth rates) form a singular eigenvalue surface with two conical points at the critical speeds
\be{ro3}
(({\rm Re}\lambda)^2+1+\Omega^2)^2-4\Omega^2=\varepsilon^2.
\ee
The slices of the surface \rf{ro3} by the planes $\varepsilon\ne0$ in the vicinity of the apexes of the cones are closed loops known as the `bubbles of instability' \cite{MK86}. Hence, for small $\varepsilon$, the Prandtl-Smith rotor is unstable at the speeds in the interval
\be{ro4}
\sqrt{1-\varepsilon}<|\Omega|<\sqrt{1+\varepsilon}.
\ee

The amount of kinetic
energy available in the rotor is usually orders of magnitude greater
than the deformational energy which any internal mode can absorb. Therefore, already small deviations from the ideal conditions yield coupling between modes that by transferring even a tiny fraction of
the rotational energy can initiate failure in the vibratory mode \cite{Cr70, Cr95}.

The conical singularities on the surfaces of frequencies and growth rates arise as unfoldings of double semi-simple eigenvalues
of the F\"oppl-von K\'arm\'an-Jeffcott rotor that exist at $\Omega=0$ and at the critical speeds. In 1986, MacKay \cite{MK86}
studied unfoldings of the semi-simple double pure imaginary and zero eigenvalues in general Hamiltonian systems and
demonstrated that the orientation of the cones depends on the Krein signature \cite{M98} of the doublets.
At $\Omega=0$ the doublets have definite Krein signature and according to MacKay the frequency cones are oriented vertically in the $(\Omega,\varepsilon,{\rm Im}\lambda)$-space that yields avoided crossings in their slices and thus stability \cite{Ki08,Ki09}. If the doublet has a mixed Krein signature then there appears a differently oriented eigenvalue cone for
the real parts of the perturbed eigenvalues that possesses bubbles of instability in its cross-sections \cite{Ki08,MK86,Ki09}.

According to \rf{ro4} the whirling instability onset
is no less than the critical speed for the mode in question. This yields a traditional for rotor dynamics
suggestion that `although Rankine
was wrong in thinking that supercritical operation was always unstable,
he was partially correct in concluding that dynamic instability would not
occur at subcritical speeds' \cite{Cr95}.

Indeed, for perturbations that preserve the Hamiltonian structure of the equations of motion of the gyroscopic system \rf{i1},
the instabilities in the subcritical speed range are prohibited by the definiteness of the Krein signature of the doublets.
However, the non-Hamiltonian perturbations caused by dissipative and non-conservative positional forces may yield dynamic instabilities at the small rotational speeds in the subcritical speed range as it happens, for example, in the problems of acoustics of friction \cite{Ki08}. Subcritical dynamic instabilities have many unusual properties partially uncovered in recent works \cite{Ki08, Ki09b, Ki09}. In the next section, we report new effects in the subcritical speed range on the example of the near-Hamiltonian Brouwer's problem.

\section{Brouwer's system in the presence of dissipative and non-conservative positional forces}

In 1976 Bottema \cite{B76} extended Brouwer's original setting by taking into account the internal and external damping
with the coefficients $\delta$ and $\nu$, respectively
\be{d1}
\ddot{\bf z}+{\bf D}\dot{\bf z}+2\Omega{\bf G}\dot{\bf z}+({\bf K}+(\Omega {\bf G})^2){\bf z}+\nu{\bf N}{\bf z}=0,
\ee
where ${\bf z}=(x,y)^T$, ${\bf D}=\diag(\delta +\nu,\delta +\nu)$, ${\bf G}={\bf J}$, ${\bf K}=\diag(k_1,k_2)$, and ${\bf N}=\Omega{\bf J}$.

Note, however, that the equation \rf{d1} appears first in a 1933 work by Smith \cite{S33} as a model
of a rotor carried by a flexible shaft in flexible bearings, where stationary (in the laboratory frame) damping coefficient $\nu$ represents the effect
of damping in bearing supports while rotating damping coefficient $\delta$ represents the effect of damping in the shaft.
These two types of damping were introduced in 1924 by  Kimball \cite{K24}
in order to explain destabilization due to damping within a rotor, studied also by Kapitsa \cite{K39}.
In a more general model of a rotating shaft by Shieh and Masur \cite{SM68} the damping matrix is allowed to be ${\bf D}=\diag(\delta_1,\delta_2)$ while the matrix of the non-conservative positional forces is simply ${\bf N}={\bf J}$.
In recent works \cite{Ki08, Ki09, Ki09b} the matrices $\bf D$ and $\bf K$ were assumed to be arbitrary symmetric while ${\bf N}={\bf J}$ with $\nu=\nu(\Omega)$ as an arbitrary smooth function of the rotation speed.

The system \rf{d1} is a particular case of a general non-conservative system with two degrees of freedom
\be{d2}
\ddot {\bf z} +{\bf B}\dot {\bf z} +{\bf A}{\bf z}=0,
\ee
where $\bf A$ and $\bf B$ are real non-symmetric matrices. Its characteristic equation is given by the Leverrier-Barnett algorithm \cite{B89}
\be{d3}
\lambda^4+{\rm tr {\bf B}}\lambda^3+({\rm tr}{\bf A}+\det {\bf B})\lambda^2+({\rm tr}{\bf A}{\rm tr}{\bf B}-{\rm tr}{\bf A}{\bf B})\lambda+\det {\bf A}=0.
\ee
The equation \rf{d3} becomes biquadratic when the following two conditions are fulfilled
\be{d4}
{\rm tr}{\bf B}=0, \quad {\rm tr}{\bf A}{\bf B}=0.
\ee
Since ${\rm tr}{\bf A}{\bf B}={\rm tr}{\bf B}{\bf A}$, these conditions are reduced to
\be{d5}
{\rm tr}{\bf B}_s=0, \quad {\rm tr}{\bf A}_s{\bf B}_s+{\rm tr}{\bf A}_a{\bf B}_a=0,
\ee
where subscripts $s$ and $a$ denote symmetric and skew-symmetric components of the matrices $\bf A$ and $\bf B$.
The conditions \rf{d5} are satisfied in particular in the case of the gyroscopic system when ${\bf B}_s=0$ and ${\bf A}_a=0$, and in the case of the circulatory system when ${\bf B}_s=0$ and ${\bf B}_a=0$.

    \begin{figure}
    \begin{center}
    \includegraphics[angle=0, width=0.8\textwidth]{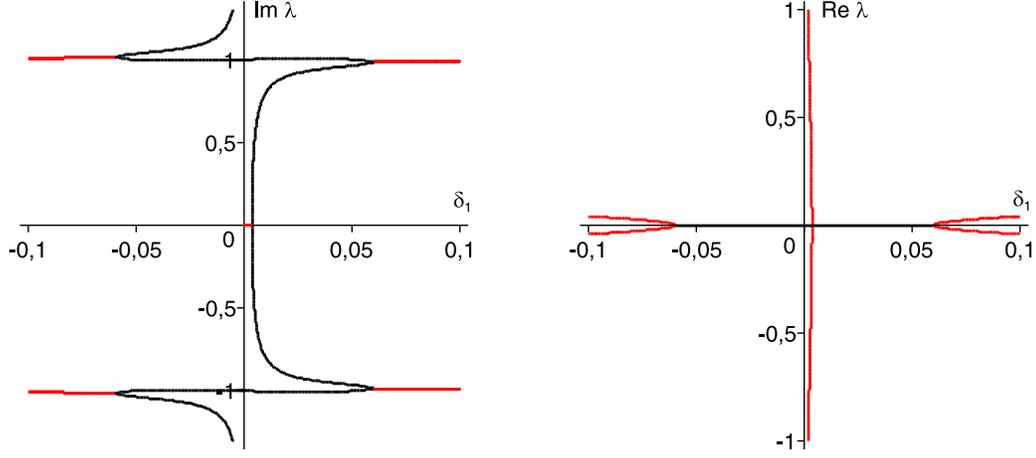}
    \end{center}
    \caption{Imaginary and real parts of the roots of the characteristic equation \rf{d6} as a function of the damping coefficient $\delta_1$ under the constraints \rf{d7} for
    $k_1=1$, $\Omega=0.03$ and $\nu=0.03$.}
    \label{fig4}
    \end{figure}

The systems, different from circulatory and gyroscopic that nevertheless satisfy conditions \rf{d5} should have some unusual properties such as an indefinite damping matrix with zero trace. Although at a first glance such a physical system looks exotic, the non-trivial example is given already by the classical Benjamin-Feir instability of the Stokes waves discussed in the previous section. Indeed, in the equation \rf{bf3} the symmetric matrix $\bf D$ is traceless and the matrix of positional forces ${\bf P}+(q{\bf J})^2$ is proportional to the unit matrix which yields fulfillment of the conditions \rf{d5}. As a consequence, the eigenvalues of the system \rf{bf3} are symmetric with respect to the imaginary axis in the complex plane which implies marginal stability if they are pure imaginary. The loss of stability of the system \rf{bf3} with the indefinite damping happens through
the collision of two imaginary eigenvalues into a double eigenvalue with the Jordan block and subsequent splitting of it into two complex ones with positive and negative real parts (Krein collision). This is not surprising in light of that the system \rf{bf3} was shown to be Hamiltonian in \cite{BD07}.

\begin{figure}
    \begin{center}
    \includegraphics[angle=0, width=0.99\textwidth]{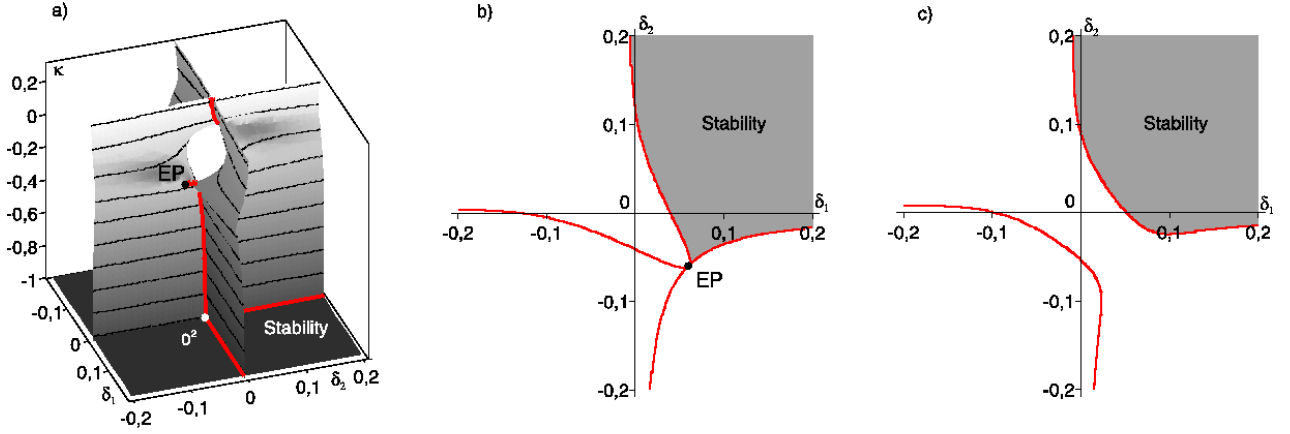}
    \end{center}
    \caption{Stability domain of the rotating shaft by Shieh and Masur for $k_1=1$, $\Omega=0.03$, $\nu=0.03$. (a) The `Viaduct' in the $(\delta_1,\delta_2,\kappa)$-space and its slices in the $(\delta_1,\delta_2)$-plane with (b) $\kappa=-0.06$ and (c) $\kappa=-0.03$.}
    \label{fig5}
    \end{figure}

The model of Shieh and Masur \cite{SM68} provides a more non-trivial example where the constraints \rf{d5} play a role.
Indeed, its characteristic equation has the form
\be{d6}
\lambda^4+(\delta_1+\delta_2)\lambda^3+
\left(\delta_1\delta_2+k_1+k_2+2\Omega^2\right)\lambda^2+
\left(k_1\delta_2+\delta_1k_2+4\Omega\nu-(\delta_1+\delta_2)\Omega^2\right)\lambda
+(\Omega^2-k_1)(\Omega^2-k_2)+\nu^2=0.
\ee
The equation \rf{d6} is biquadratic in case when
\be{d7}
\delta_1+\delta_2=0, \quad \kappa=-\frac{4\Omega\nu }{\delta_1},
\ee
with $\kappa=k_2-k_1$. If $k_1>\Omega^2$ and $\nu>0$ then all the roots of the equation \rf{d6} are pure imaginary when
\be{d8}
-2\Omega\le\delta_1<0,\quad\frac{4\Omega\nu(k_1-\Omega^2)}{\nu^2+(k_1-\Omega^2)^2}<\delta_1\le2\Omega.
\ee
In Fig.~\ref{fig4} the pure imaginary eigenvalues are shown by black lines as functions of the damping parameter $\delta_1$.
At
\be{d8a}
\delta_1=\delta_d:=\frac{4\Omega\nu(k_1-\Omega^2)}{\nu^2+(k_1-\Omega^2)^2}, \quad \kappa=\kappa_d:=-k_1+\Omega^2-\frac{\nu^2}{k_1-\Omega^2}
\ee
there exist a double zero eigenvalue with the Jordan block $(0^2)$,
see Fig.~\ref{fig4} and Fig.~\ref{fig5}(a). In the interval $0<\delta_1<\delta_d$ there exist one positive and one negative real eigenvalue.
In Fig.~\ref{fig4} the eigenvalues with non-zero real parts are shown in red.
In the $(\delta_1,\delta_2,\kappa)$-space the \textit{exceptional points} (EPs)
\be{d9}
(-2\Omega, 2\Omega , 2\nu), \quad (2\Omega, -2\Omega, -2\nu)
\ee
correspond to the double pure imaginary eigenvalues with the Jordan block
\be{d10}
\lambda_{-2\Omega}=\pm i \sqrt{{k_1}-\Omega^2+{\nu}}, \quad \lambda_{2\Omega}=\pm i \sqrt{{k_1}-\Omega^2- {\nu}},
\ee
for $\delta_1=-2\Omega$ and $\delta_1=2\Omega$, respectively.

We see in Fig.~\ref{fig4} that changing the damping parameter $\delta_1$ we migrate from the marginal stability domain to that of flutter instability by means of the Krein collision of the two simple pure imaginary eigenvalues as it happens in gyroscopic or circulatory systems without dissipation. It is remarkable that such a behavior of eigenvalues is observed in the gyroscopic system in the presence of dissipative and non-conservative positional forces. The existence of pure imaginary spectrum in the non-conservative dissipative systems satisfying the constraints \rf{d5} reminds the similar phenomenon in PT-symmetric non-Hermitian systems that can possess pure real spectrum \cite{BHH07}.

Let us now establish how in the $(\delta_1,\delta_2,\kappa)$-space the domain of marginal stability given by the expressions \rf{d7} and \rf{d8} is connected to the domain of asymptotic stability of the non-constrained equation \rf{d6}. Writing the
${\rm Li\acute{e}nard}$ and Chipart \cite{K06} conditions for asymptotic stability of the polynomial \rf{d6} we find
\ba{d11}
p_1:=\delta_1+\delta_2&>&0,\nn \\
p_2:=\delta_1\delta_2+k_1+k_2+2\Omega^2&>&0,\nn \\
p_4:=(\Omega^2-k_1)(\Omega^2-k_2)+\nu^2&>&0,\nn \\
H_3:=(\delta_1+\delta_2)(\delta_1\delta_2+k_1+k_2+2\Omega^2)(k_1\delta_2+\delta_1k_2+4\Omega\nu-(\delta_1+\delta_2)\Omega^2)& &\nn\\
-(\delta_1+\delta_2)^2((\Omega^2-k_1)(\Omega^2-k_2)+\nu^2 )-(k_1\delta_2+\delta_1k_2+4\Omega\nu-(\delta_1+\delta_2)\Omega^2)^2&>&0.
\ea

The surfaces $p_4=0$ and $H_3=0$ are plotted in Fig.~\ref{fig5}(a). The former is simply a horizontal plane that passes through the point of the double zero eigenvalue $(0^2)$ with the coordinates $(\delta_d,-\delta_d,\kappa_d)$ and thus bounds the stability domain from below. The surface $H_3=0$ is singular because it has self-intersections along the portions of the hyperbolic curves \rf{d7} selected by the inequalities \rf{d8}. The curve of self-intersection that corresponds to $\kappa>0$ ends up at the EP with the double pure imaginary eigenvalue $\lambda_{-2\Omega}$. Another curve of self-intersection has at its ends the EP
with the double pure imaginary eigenvalue $\lambda_{2\Omega}$ and the point of the double zero eigenvalue $(0^2)$. In Fig.~\ref{fig5}(a) the curves of self-intersection are shown in red and the EP and $0^2$ are marked by the black and white circles, respectively. At the point $0^2$ the surfaces $p_4=0$ and $H_3=0$ intersect each other forming a trihedral angle singularity of the stability boundary with its edges depicted by red lines in Fig.~\ref{fig5}(a). The surface $H_3=0$ is symmetric with respect to the plane $p_1=0$. Thus, a part of it that belongs to the subspace $p_1>0$ bounds the domain of asymptotic stability.

At the EPs, the boundary of the asymptotic stability domain has singular points that are locally equivalent to the {\em Whitney umbrella} singularity \cite{B56, Ar71, GKL1990, HK10, KV10}. Between the two EPs the surface $H_3=0$ has an opening around the origin that separates its two sheets. This window allows the flutter instability to exist in the vicinity of the origin for small damping coefficients and small separation of the stiffness coefficients $\kappa$. In \cite{Ki09} an eigenvalue surface of the same shape was named \textit{the Viaduct}.

In Fig.~\ref{fig5}(b) a cross-section of the surface $H_3=0$ by the horizontal plane that passes through the lower exceptional point is shown. The domain in grey indicates the area of asymptotic stability. Its boundary has a cuspidal point singularity at the EP. Although the  very singular shape of the planar stability domain is typical in the vicinity of the EP with the pure imaginary double eigenvalue with the Jordan block \cite{KV10}, the unusual feature is the location of the EP that corresponds to non-vanishing damping coefficients, Fig.~\ref{fig5}(b).

Indeed, in the known studies, the system with vanishing dissipation was either gyroscopic (Hamiltonian) as in the works \cite{K06, KV10, K07a, K07b, MK91, KM07, Sa08} or it was circulatory (reversible) as in the works \cite{B56, KV10, K07a, Zi52}. The undamped system in such problems has a domain of marginal stability bounded by the exceptional points
that correspond to double pure imaginary eigenvalues. This is not the case in the Brouwer's problem with the dissipative and non-conservative forces by Shieh and Masur \cite{SM68}. According to the theorems of Bottema \cite{B55} and Lakhadanov \cite{L75} the undamped gyroscopic system with non-conservative positional forces is generically unstable. By examining the slices of the surface $H_3=0$ at various values of $\kappa$ one can see that the origin is indeed always unstable, Fig.~\ref{fig5}(b,c). At $\kappa=0$ the origin is unstable in the presence of the non-conservative positional forces even when the rotation is absent $(\Omega=0)$ according to the theorem of Merkin \cite{KM07}. Contrary to the situation known as the \textit{Ziegler's destabilization paradox} \cite{Zi52}, in the Brouwer-Shieh-Masur model the tending of the damping coefficients to zero along a path in the $(\delta_1,\delta_2)$-plane cannot lead to the set of pure imaginary spectrum of the undamped system because in this model such a set corresponds to the non-vanishing damping coefficients.

On the other hand, the lower bound $\kappa_d$ on the difference of the coefficients $k_{1,2}$ given by the equation \rf{d8a} depends on the magnitude of the non-conservative positional force $\nu$. If $\nu=0$, the heavy particle on a rotating saddle is destabilized by the arbitrary small internal damping, because in this case $\kappa_d=-k_1+\Omega^2$ and the stability condition $\kappa>\kappa_d$ reduces to $\kappa_2>\Omega^2$, in full agreement with \cite{B76}. Nevertheless, with the increase of $\nu$ the critical difference of $k_1$ and $k_2$ takes bigger negative values so that for $\nu^2>\Omega^2(k_1-\Omega^2)$ asymptotic stability is possible also for the coefficient $k_2<0$ while $k_1$ was assumed to be positive.

Therefore, the Brouwer-Shieh-Masur model provides a non-trivial example of a gyroscopic system
that can have all its eigenvalues pure imaginary in the presence of dissipative and non-conservative forces. The corresponding set
in the
three-dimensional space of damping coefficients and the difference between the curvatures (stiffness coefficients) is a space curve that
ends up at the exceptional points of double pure imaginary eigenvalues with the Jordan block. The set forms singularities of the boundary to the domain of asymptotic stability of the Brouwer-Shieh-Masur model such as Whitney umbrellas and dihedral and trihedral angles. The highly non-trivial shape of the discovered stability boundary illustrates the peculiarities of destabilization of a rotor in the subcritical speed range as well as the stabilization of a heavy particle on a rotating saddle for high rotating velocities by non-conservative positional forces in their interplay with the dissipative, gyroscopic and potential ones.

\section{Conclusion}

The Brouwer's problem on a heavy particle in a rotating vessel is a paradigmatic model
for studying instabilities in diverse areas of physics. Brouwer-like systems of equations naturally arise for instance in accelerator physics,
crystal optics, theory of Stokes waves in deep water, as well as in classical rotor dynamics.
Despite a long history, the model of Brouwer was not completely understood until now. We established new results on stability
in the Brouwer's problem that include discovery of the Swallowtail singularity on the stability boundary
of the undamped Brouwer's equations. This singularity was overlooked both by Brouwer and by Bottema who re-visited Brouwer's work in 1976. Moreover, we studied Brouwer's equations in the presence of dissipative and non-conservative forces and found a new type of the asymptotic stability domain in the three-dimensional space of parameters. It turns out that the stability boundary is a surface with two Whitney umbrella singularities, the `handles' of which correspond to systems with the pure imaginary spectrum despite the presence of damping and non-conservative positional forces. We established the necessary conditions for a non-conservative system to have a pure imaginary spectrum and found that the linearized equations describing the onset of the Benjamin-Feir instability satisfy them. As a consequence, the Benjamin-Feir instability was interpreted as destabilization of a gyroscopic system by indefinite damping.

To summarize, the Brouwer's model describes two common situations: a heavy particle in the narrow concave rotating vessel is difficult to destabilize; that one in the wide rotating dish or on a rotating saddle is difficult to stabilize. The former  is common in the modern problems of low-speed rotor dynamics related to squealing brakes or singing glasses of the glass harmonica. The latter is typical for the problems of particle confinement in atomic and accelerator physics as well as for high-speed rotor dynamics. In both situations the destabilization and stabilization mechanisms exist that involve dissipative and other non-conservative forces.
The very possibility to create (in)stability in such circumstances where it conventionally is not expected makes such problems
a reach source for new development and insight in stability theory. The Brouwer's problem on a heavy particle in a rotating vessel sharply highlights some crucial effects that arise in the field of dissipation-induced instabilities.

\section*{Acknowledgements}
The work has been supported by the Alexander von Humboldt Foundation.

\end{document}